\RequirePackage[undo-recent-deprecations]{expl3}
\documentclass[a4paper,fleqn]{cas-sc}
\usepackage{soul}
\usepackage[sort&compress,numbers]{natbib}
\usepackage{lineno}
\usepackage{setspace}
\usepackage{multirow}
\usepackage{lipsum}

\usepackage{cuted}

\def\tsc#1{\csdef{#1}{\textsc{\lowercase{#1}}\xspace}}
\tsc{WGM}
\tsc{QE}
\tsc{EP}
\tsc{PMS}
\tsc{BEC}
\tsc{DE}

\begin{document}


\let\WriteBookmarks\relax
\def\floatpagepagefraction{1}
\def\textpagefraction{.001}
\shorttitle{Mechanical Properties of \texorpdfstring{MoX$_2$}{} (X = S, Se, Te) Membranes}
\shortauthors{Pereira J\'unior \textit{et~al}.}

\title [mode = title]{On the Elastic Properties and Fracture Patterns of \texorpdfstring{MoX$_2$}{} (X = S, Se, Te) Membranes: A Reactive Molecular Dynamics Study}

\author[1]{M. L. Pereira J\'unior}
\author[1]{C. M. Viana de Ara\'ujo}
\author[2]{J. M. de Sousa}
\author[3]{R. T. de Sousa J\'unior}
\author[1]{L. F. Roncaratti J\'unior}
\author[3]{W. F. Giozza}
\author[1]{L. A. Ribeiro J\'unior}
\cormark[1]
\ead{ribeirojr@unb.br}

\address[1]{Institute of Physics, University of Bras\'ilia, Bras\'ilia, 70910-900, Brazil.}

\address[2]{Federal Institute of Education, Science and Technology of Piau\'i, S\~ao Raimundo Nonato, Piau\'i, Brazil.}

\address[3]{Department of Electrical Engineering, University of Bras\'{i}lia 70919-970, Brazil.}


\begin{abstract}
We carried out fully-atomistic reactive molecular dynamics simulations to study the elastic properties and fracture patterns of transition metal dichalcogenide (TMD) MoX$_{2}$ (X=S,Se,Te) membranes, in their 2H and 1T phases, within the framework of the Stillinger-Weber potential. Results showed that the fracture mechanism of these membranes occurs through a fast crack propagation followed by their abrupt rupture into moieties. As a general trend, the translated arrangement of the chalcogen atoms in the 1T phase contributes to diminishing their structural stability when contrasted with the 2H one. Among the TMDs studied here, 2H-MoSe$_{2}$ has higher tensile strength (25.98 GPa). 
\end{abstract}


\maketitle
\doublespacing

\section{Introduction}

Transition metal dichalcogenide (TMD) monolayers are atomically thin semiconductors that belong to the family of $2D$ nanosheets \cite{manzeli20172d,chhowalla2013chemistry}. They present an MX$_{2}$ arrangement, where M is a transition metal, and X is a chalcogen. The combination of chalcogen (e.g., S, Se, or Te) and transition metal atoms (typically Mo, W, Nb, Re, Ni, or V) yields more than 40 different materials \cite{tan2015two,chhowalla2015two}. Each monolayer has a thickness of $6-7$ \r{A} and is hexagonally-packed by transition metal atoms sandwiched between two layers of chalcogen atoms \cite{tan2015two}. TMDs are graphene cognate and possible to synthesize by applying the same chemical methods usually employed in producing the latter \cite{xu2013graphene,butler2013progress}. These materials have received much attention in the fields of biomedicine \cite{qian2015two,chen2015two}, optoelectronics \cite{wang2012electronics,wilson1969transition}, and energy conversion and storage \cite{eda2011photoluminescence,yun2018three}. Particularly, MoS$_2$ and MoTe$_2$ monolayers --- direct bandgap semiconductors with bandgaps about 1.9 eV \cite{mak2010atomically} and 1.0 eV \cite{ruppert2014optical}, respectively --- have emerged as promising candidates in replacing gapless graphene to develop novel applications in which semiconducting materials are desired \cite{jariwala2014emerging}. MoSe$_2$, in turn, is an indirect bandgap semiconductor (with a bandgap about 1.58 eV \cite{zhang2014direct}) that has also been widely employed in developing new applications in flat electronics \cite{eftekhari2017molybdenum,eda2013two}. To further explore the potential of these TMDs species in boosting new advances in the research fields mentioned above, their mechanical properties should be deeply understood. 

TMD nanostructures have three different structural arrangements, named 2H, 1T, and 1T' \cite{zhang2018novel}. 2H and 1T refer to the hexagonal and trigonal structures, respectively. The 1T' phase is a distorted form of 1T. Significant theoretical and experimental efforts have been employed in understanding the mechanical properties of layered MoS$_2$ \cite{alireza_JPCL,andres_AM01,andres_AM02,andres_NL,cooper_PRB,jiang_APL,jiang_JAP,kang_PCCP,khan_AMSE,liang_PRB,liu_NL,manzeli_NL,yazev_MT,mortazavi_PCCP,miro_ACI}, MoSe$_2$ \cite{frisenda_NPJ2DMAT,jaques_MRSADV,jiang_JMCA,iguiniz2019revisiting}, and MoTe$_2$ \cite{johari_ACSNANO,kumar_PB,mortazaviEM,ruppert_NL,may2013reinforcement,rano2020ab,sun2019elastic} on both 2H and 1T forms. In these investigations, it was experimentally studied few layers (5-25) of these TMDs species, and Young's modulus obtain were approximately 330 Gpa \cite{andres_AM01}, 117 Gpa \cite{iguiniz2019revisiting}, and 110 Gpa \cite{may2013reinforcement} for MoS$_2$, MoSe$_2$, and MoTe$_2$, respectively. By using density functional theory and reactive molecular dynamic simulations, theoretical studies have predicted Yong's modulus values for single-layer MoS$_2$, MoSe$_2$, and MoTe$_2$ ranging in the intervals 170-250 Gpa \cite{,alireza_JPCL,cooper_PRB,mortazavi_PCCP}, 165-185 Gpa \cite{deng_PE, iguiniz_AM}, and 60-115 Gpa \cite{mortazaviEM,rano2020ab,sun2019elastic}, respectively. These works promoted substantial advances in understanding the mechanical properties of TMDs. However, an overall description of their elastic properties and fracture dynamics is still missing.         

Herein, we carried out extensive fully-atomistic reactive molecular dynamics simulations to study the elastic properties and fracture dynamics of MoX$_{2}$ (X=S,Se,Te) membranes in their 2H and 1T phases. The elastic properties were obtained from the stress-strain relationship. Only recently, 1T phases of these materials were experimentally realized \cite{zhang2018novel}. In this sense, a detailed description of the mechanical properties of these nanostructures considering both 2H and 1T phases is highly attractive.    

\section{Methodology}

\begin{figure*}[pos=ht]
	\centering
	\includegraphics[width=0.95\linewidth]{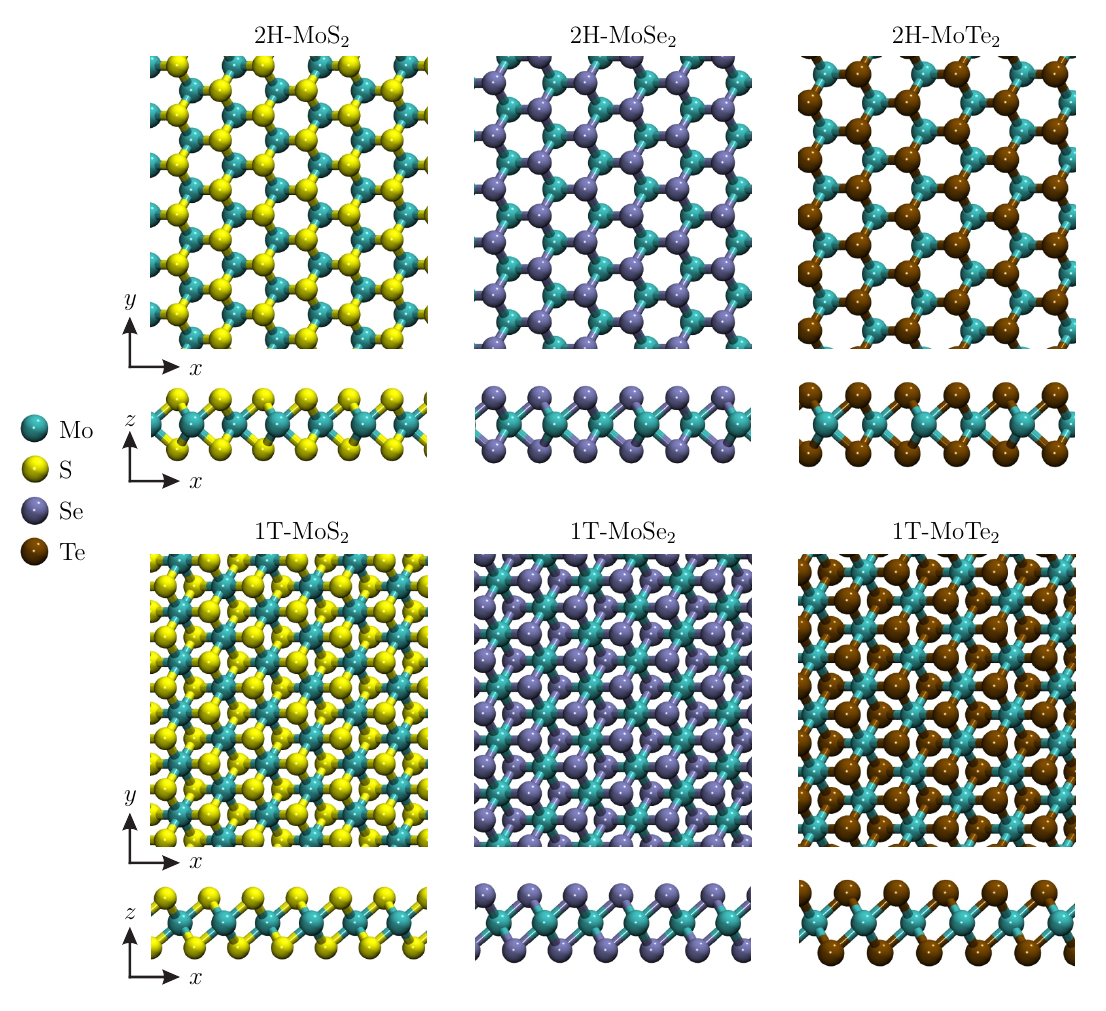}
	\caption{Schematic representation of the model TMDs monolayers in their 2H and 1T phases.}
	\label{fig:structures}
\end{figure*}

We performed fully-atomistic molecular dynamics (MD) simulations using the Stillinger-Weber (SW) \cite{jiang_JAP,jiang_APL} potential as implemented by LAMMPS \cite{plimpton1995fast}. Figure \ref{fig:structures} illustrates the model TMDs monolayers studied here in their 2H and 1T phases. The left, middle, and bottom panes illustrate the MoS$_2$, MoSe$_2$, and MoTe$_2$ monolayers, respectively, in the H (top panels) and T phases (bottom panels). Their atomistic structure contains 3456, 3348, and 2688 atoms, respectively, and they were built intended in yielding 2D membranes with dimensions of about $100\times 100$ \AA${^2}$, with periodic boundary conditions.   

The equations of motion were solved using the velocity-Verlet integrator with a time-step of $0.1$ fs. The tensile stress was considered in the system by applying a uniaxial strain along the periodic $x$ and $y$ directions, using an engineering strain rate of $10^{-5}$ fs$^{-1}$. The TMD membranes were stressed up to their complete rupture. To eliminate any residual stress within the membranes, they were equilibrated within an NPT ensemble at constant temperatures (1K and 300K) and null pressures using the Nos\'e-Hoover thermostat during 200 ps. By adopting this simulation protocol, Young’s modulus ($Y$), Fracture Strain ($FS$), and Ultimate Strength ($US$) are the elastic properties derived from the stress-strain curves. The fracture dynamics, in turn, are studied through MD snapshots and the von Mises stress (VM) per-atom values, calculated at every 100 fs \cite{pereirajr_FC}. The VM values provide useful local structural information on the fracture mechanism, once they can determine the region from which the structure has started to yield the fractured lattice. The MD snapshots and trajectories were obtained by using free visualization and analysis software VMD \cite{HUMPHREY199633}. 

\section{Results}

\begin{figure*}[pos=ht]
	\centering
	\includegraphics[width=0.95\linewidth]{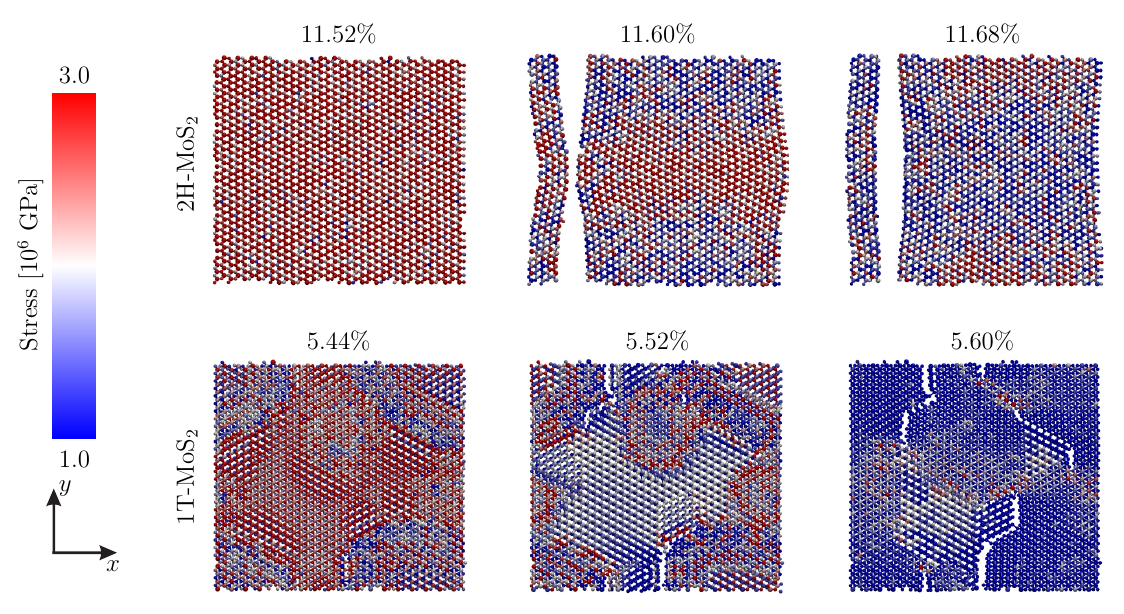}
	\caption{Representative MD snapshots of the fracture dynamics for the 2H-MoS$_2$ (top sequence of panels) and 1T-MoS$_2$ (bottom sequence of panels) monolayers at 300K and considering a uniaxial strain applied along the $x$-direction.}
	\label{fig:mos2x}
\end{figure*}

We begin our discussions by showing representative MD snapshots of the fracture dynamics for the 2H-MoS$_2$ (top sequence of panels) and 1T-MoS$_2$ (bottom sequence of panels) monolayers at 300K and considering a uniaxial strain applied along the $x$-direction, as shown in Figure \ref{fig:mos2x}. In the 2H-MoS$_2$ case, one can note an abrupt rupture followed by a fast propagation of the fracture along the $y-direction$ is accomplished at 11.60\% of strain. The membrane is considered fractured at 11.68\% of strain, once the atoms in the edges of the two fractures moieties are not interacting. A different fracture trend is realized for a 1T-MoS$_2$ membrane. The very first striking outcome obtained here, when contrasting the fracture dynamics of 2H-MoS$_2$ and 1T-MoS$_2$, is the considerably higher degree of fragility of the latter case. In Figure \ref{fig:mos2x}, one can observe that the critical strain for the beginning of the fracture in the 1T-MoS$_2$ (5.44\%) is almost two times smaller than the one for 2H-MoS$_2$. Another clear trend showed in this figure is that the fracture dynamics of 1T-MoS$_2$ leads to a brittle lattice structure. This rupture trend is different from the one obtained for the 2H-MoS$_2$ case, in which two well concise MoS$_2$ fragments were produced as a final stage of the fracture process. This brittle signature for the 1T-MoS$_2$ case is obtained for 5.60\% of strain. These results suggest that the translated arrangement of the chalcogen atoms in the 1T phase is crucial in diminishing the structural stability of TMDs.                      

\begin{figure*}[pos=ht]
\centering
\includegraphics[width=0.95\linewidth]{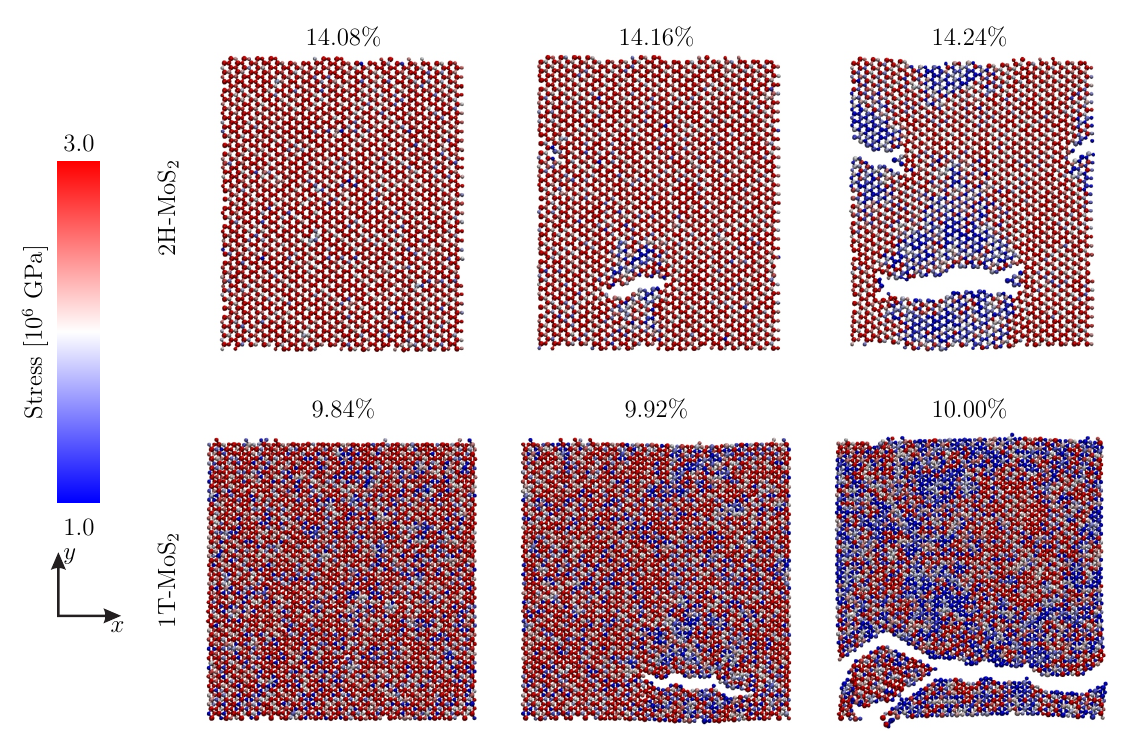}
\caption{Representative MD snapshots of the fracture dynamics for the 1T-MoS$_2$ (top sequence of panels) and 1T-MoS$_2$ (bottom sequence of panels) monolayers at 300K and considering a uniaxial strain applied along the $y$-direction.}
\label{fig:mos2y}
\end{figure*}

An interesting result arises when a uniaxial strain is applied along the $y$-direction, as depicted in Figure \ref{fig:mos2y}. This figure shows the cases related to Figure \ref{fig:mos2x}. When the tensile stretching is applied in the y-direction, the critical strain to realize the beginning of the fracture of 2H-MoS$_2$ and 1T-MoS$_2$ membranes is considerably higher than the ones presented in Figure \ref{fig:mos2x}). The difference between the fracture strains for these species is smaller when the stretching is applied along the $y$-direction. As illustrated in Figure \ref{fig:mos2y}, the fracture (critical) strains for the beginning of the rupture are 14.08\% and 9.84\% for the 2H-MoS$_2$ and 1T-MoS$_2$ membranes, respectively. After that critical value, the crack propagation takes place for 14.16\% and 9.92\% for the 2H-MoS$_2$ and 1T-MoS$_2$ cases, respectively. Interestingly, the brittle trend for the 1T-MoS$_2$ fracture, obtained for the $y$-direction stretching, not occurs when it comes to the $y$-direction stretching. After 14.24\% and 10.0\%, a fast crack propagation occurs for the 2H-MoS$_2$ and 1T-MoS$_2$ monolayers, respectively. 

\begin{figure*}[pos=ht]
\centering
\includegraphics[width=0.95\linewidth]{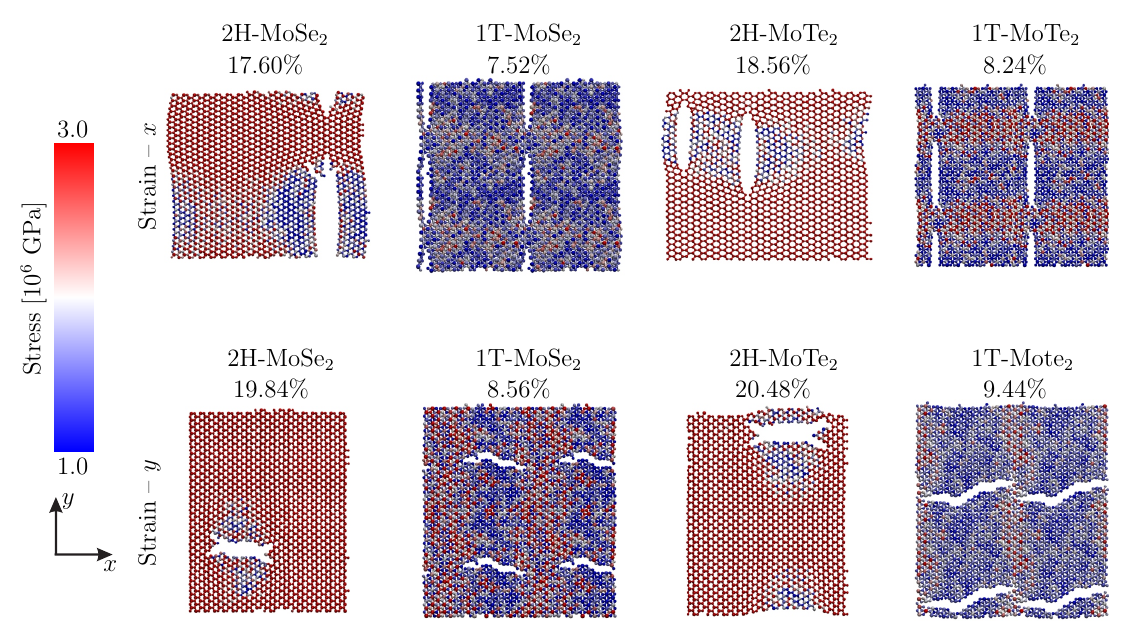}
\caption{Representative MD snapshots of the fracture dynamics for the 2H/1T-MoSe$_2$ and 2H/1T-MoTe$_2$ monolayers at 300K. The top and bottom sequence of panels refer to the simulations considering a uniaxial strain applied along the $x$ and $y$ directions, respectively.}
\label{fig:mose2mote2}
\end{figure*}

Figure \ref{fig:mose2mote2} illustrates the representative MD snapshots for the 2H/1T-MoSe$_{2}$ and 2H/1T-MoTe$_{2}$ membranes. For the sake of convenience, we presented just the snapshots that show the crack propagation and the fractured form of these TMD species. The top and bottom panels depict the results when the uniaxial strain is applied along the $x$ and $y$ directions, respectively. In the top panels, one can note that the critical strain for the membrane rupture is 17.60\%, 7.52\%, 18.56\%, and, 8.24\% for the 2H-MoSe$_{2}$, 1T-MoSe$_{2}$, 2H-MoTe$_{2}$, and 1T-MoTe$_{2}$, respectively. In the bottom panels, we can observe that the critical strain for the membrane rupture is 19.84\%, 8.56\%, 20.48\%, and, 9.44\% for the 2H-MoSe$_{2}$, 1T-MoSe$_{2}$, 2H-MoTe$_{2}$, and 1T-MoTe$_{2}$, respectively. A comparison of the tensile strength among the TMDs studied here is presented below with Table \ref{table:elasprop}. As for the MoS$_2$ cases, in the MoSe$_{2}$ and MoTe$_{2}$ cases, the fracture propagation undergoes in the direction opposite to the stretching. It is worthwhile to stress that both MoSe$_{2}$ and MoTe$_{2}$ present a fracture mechanism defined by a fast crack propagation followed by an abrupt rupture of the membranes into parts with a good degree of integrity (i.e., no brittle structures were observed). Theses results suggest that MoSe$_2$ and MoTe$_2$ monolayers may present greater structural stability than the MoS$_2$ ones.      

\begin{figure*}[pos=ht]
\centering
\includegraphics[width=0.95\linewidth]{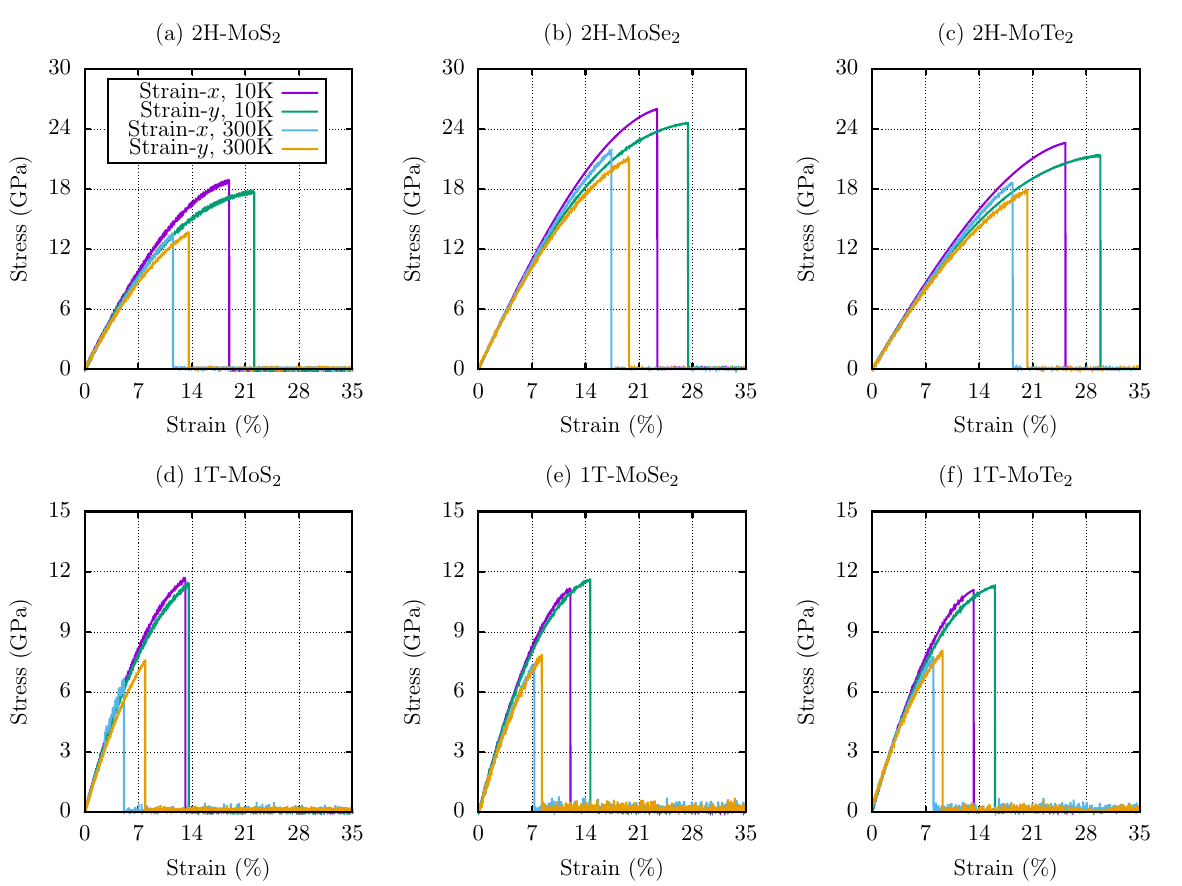}
\caption{Calculated stress-strain curves for all TMD membranes studied here when subjected to 10K and 300K, considering a uniaxial strain applied in both $x$ and $y$ directions. Figures \ref{fig:curves}(a-d), \ref{fig:curves}(b-e), and \ref{fig:curves}(c-f) are describing the stress-strain relationship for the 2H/1T-MoS$_2$, 2H/1T-MoSe$_2$, and 2H/1T-MoTe$_2$ membranes, respectively.}

\label{fig:curves}
\end{figure*}

Finally, we present the elastic properties obtained from the simulations discussed above. These properties are Young's modulus ($Y_{M}$, in units of GPa), fracture strain ($FS (\%)$), and the maximum stress $\sigma_{US}$ (which is called Ultimate Tensile Strength US (GPa)). They were estimated considering stretching regimes before mechanical failure (fracture) of the TMD membranes. These stretching regimes can be inferred from Figure \ref{fig:curves}, which illustrates the calculated stress-strain curves for all TMD membranes when subjected to 10K and 300K, considering a uniaxial strain applied in both $x$ and $y$ directions. Figures \ref{fig:curves}(a-d), \ref{fig:curves}(b-e), and \ref{fig:curves}(c-f) are describing the stress-strain relationship for the 2H/1T-MoS$_2$, 2H/1T-MoSe$_2$, and 2H/1T-MoTe$_2$ membranes, respectively. Table \ref{table:elasprop} present a summary of the mechanical properties of the TMD monolayers studied in this work. In our simulation protocol, these monolayers were stretched at a constant rate until their total rupture. The stress-strain curves shown the two following common regions: a quasi-linear elastic region that is observed up to the ultimate strength value and a region of null stress (after a critical fracture strain) in which the TMD membranes ultimately break. In Figure \ref{fig:curves}, one can see that the $US$ values are slightly higher for the cases in which the tensile stretching was applied in the $y$-direction. This trend occurs since the bond angle variations in $x-$ and $y-$direction are different, and they govern the fracture strain. The fracture strains range from 5.44\% (1T-MoS$_2$ at 300K) up to 29.86\% (2H-MoTe$_2$ at 10K). As expected, increasing the temperature to 300K, there are a decrease in the critical tensile strain (fracture strain) values for all TMD membranes (see \ref{table:elasprop}). The highest Young's Modulus was obtained for 2H-MoSe$_2$ monolayer at 10K (154.65 GPa). The TMD of the higher tensile strength (highest ultimate stress value) is the 2H-MoSe$_{2}$ membrane at 10K (25.98 GPa). As discussed above, generally, the translated arrangement of the chalcogen atoms in the 1T phase can contribute to diminishing their structural stability when compared with TMD membranes in the 2H phase. Importantly, Table \ref{table:elasprop} summarizes the elastic properties ($Y_M$, $FS$, and $US$) that were obtained by fitting the stress-strain curves for the TMD monolayers investigated here.

	\begin{table*}[]
		\begin{tabular}{clccclccc}
			\hline
			\multicolumn{1}{c}{\multirow{3}{*}{Temperature [K]}} & \multicolumn{1}{l}{} &                                                                                                      \multicolumn{7}{c}{2H-MoS$_2$}                                                                                                       \\ \cline{3-9}
			                  \multicolumn{1}{c}{}                   & \multicolumn{1}{l}{} &                                     \multicolumn{3}{c}{ Strain-$x$}                                      & \multicolumn{1}{l}{} &                                      \multicolumn{3}{c}{Strain-$y$}                                      \\ \cline{3-5}\cline{7-9}
			                  \multicolumn{1}{c}{}                   & \multicolumn{1}{l}{} & \multicolumn{1}{c}{$Y_M$ {[}GPa{]}} & \multicolumn{1}{c}{FS {[}\%{]}} & \multicolumn{1}{c}{US {[}GPa{]}} & \multicolumn{1}{l}{} & \multicolumn{1}{c}{$Y_M$ {[}GPa{]}} & \multicolumn{1}{c}{FS {[}\%{]}} & \multicolumn{1}{c}{US {[}GPa{]}} \\ \cline{1-1}\cline{3-5}\cline{7-9}
			                          10 K                            &                      &               144.92                &              18.89              &              18.81               &                      &               139.43                &              22.16              &              17.76               \\
			                \multicolumn{1}{c}{300 K}                 & \multicolumn{1}{l}{} &     \multicolumn{1}{c}{135.13}      &    \multicolumn{1}{c}{11.54}    &    \multicolumn{1}{c}{13.57}     & \multicolumn{1}{l}{} &     \multicolumn{1}{c}{130.13}      &    \multicolumn{1}{c}{13.60}    &    \multicolumn{1}{c}{13.80}
		\end{tabular}
		
		\begin{tabular}{clccclccc}
			\hline
			\multicolumn{1}{c}{\multirow{3}{*}{Temperature [K]}} & \multicolumn{1}{l}{} &                                                                                                      \multicolumn{7}{c}{2H-MoSe$_2$}                                                                                                      \\ \cline{3-9}
			                  \multicolumn{1}{c}{}                   & \multicolumn{1}{l}{} &                                     \multicolumn{3}{c}{ Strain-$x$}                                      & \multicolumn{1}{l}{} &                                      \multicolumn{3}{c}{Strain-$y$}                                      \\ \cline{3-5}\cline{7-9}
			                  \multicolumn{1}{c}{}                   & \multicolumn{1}{l}{} & \multicolumn{1}{c}{$Y_M$ {[}GPa{]}} & \multicolumn{1}{c}{FS {[}\%{]}} & \multicolumn{1}{c}{US {[}GPa{]}} & \multicolumn{1}{l}{} & \multicolumn{1}{c}{$Y_M$ {[}GPa{]}} & \multicolumn{1}{c}{FS {[}\%{]}} & \multicolumn{1}{c}{US {[}GPa{]}} \\ \cline{1-1}\cline{3-5}\cline{7-9}
			                          10 K                            &                      &               159.63                &              23.39              &              25.98               &                      &               154.65                &              27.41              &              24.59               \\
			                \multicolumn{1}{c}{300 K}                 & \multicolumn{1}{l}{} &     \multicolumn{1}{c}{153.81}      &    \multicolumn{1}{c}{17.39}    &    \multicolumn{1}{c}{21.88}     & \multicolumn{1}{l}{} &     \multicolumn{1}{c}{148.80}      &    \multicolumn{1}{c}{19.68}    &    \multicolumn{1}{c}{21.16}
		\end{tabular}
		
		\begin{tabular}{clccclccc}
			\hline
			\multicolumn{1}{c}{\multirow{3}{*}{Temperature [K]}} & \multicolumn{1}{l}{} &                                                                                                      \multicolumn{7}{c}{2H-MoTe$_2$}                                                                                                      \\ \cline{3-9}
			                  \multicolumn{1}{c}{}                   & \multicolumn{1}{l}{} &                                     \multicolumn{3}{c}{ Strain-$x$}                                      & \multicolumn{1}{l}{} &                                      \multicolumn{3}{c}{Strain-$y$}                                      \\ \cline{3-5}\cline{7-9}
			                  \multicolumn{1}{c}{}                   & \multicolumn{1}{l}{} & \multicolumn{1}{c}{$Y_M$ {[}GPa{]}} & \multicolumn{1}{c}{FS {[}\%{]}} & \multicolumn{1}{c}{US {[}GPa{]}} & \multicolumn{1}{l}{} & \multicolumn{1}{c}{$Y_M$ {[}GPa{]}} & \multicolumn{1}{c}{FS {[}\%{]}} & \multicolumn{1}{c}{US {[}GPa{]}} \\ \cline{1-1}\cline{3-5}\cline{7-9}
			                          10 K                            &                      &               125.21                &              25.29              &              22.61               &                      &               121.54                &              29.86              &              21.37               \\
			                \multicolumn{1}{c}{300 K}                 & \multicolumn{1}{l}{} &     \multicolumn{1}{c}{121.46}      &    \multicolumn{1}{c}{18.39}    &    \multicolumn{1}{c}{18.70}     & \multicolumn{1}{l}{} &     \multicolumn{1}{c}{117.63}      &    \multicolumn{1}{c}{20.30}    &    \multicolumn{1}{c}{17.98}
		\end{tabular}
	
		\begin{tabular}{clccclccc}
			\hline
			\multicolumn{1}{c}{\multirow{3}{*}{Temperature [K]}} & \multicolumn{1}{l}{} &                                                                                                      \multicolumn{7}{c}{1T-MoS$_2$}                                                                                                       \\ \cline{3-9}
			                  \multicolumn{1}{c}{}                   & \multicolumn{1}{l}{} &                                     \multicolumn{3}{c}{ Strain-$x$}                                      & \multicolumn{1}{l}{} &                                      \multicolumn{3}{c}{Strain-$y$}                                      \\ \cline{3-5}\cline{7-9}
			                  \multicolumn{1}{c}{}                   & \multicolumn{1}{l}{} & \multicolumn{1}{c}{$Y_M$ {[}GPa{]}} & \multicolumn{1}{c}{FS {[}\%{]}} & \multicolumn{1}{c}{US {[}GPa{]}} & \multicolumn{1}{l}{} & \multicolumn{1}{c}{$Y_M$ {[}GPa{]}} & \multicolumn{1}{c}{FS {[}\%{]}} & \multicolumn{1}{c}{US {[}GPa{]}} \\ \cline{1-1}\cline{3-5}\cline{7-9}
			                          10 K                            &                      &               123.98                &              13.16              &              11.66               &                      &               120.42                &              13.60              &              11.42               \\
			                \multicolumn{1}{c}{300 K}                 & \multicolumn{1}{l}{} &     \multicolumn{1}{c}{132.65}      &    \multicolumn{1}{c}{5.11}     &     \multicolumn{1}{c}{6.91}     & \multicolumn{1}{l}{} &     \multicolumn{1}{c}{110.00}      &    \multicolumn{1}{c}{7.90}     &     \multicolumn{1}{c}{7.62}
		\end{tabular}

		\begin{tabular}{clccclccc}
			\hline
			\multicolumn{1}{c}{\multirow{3}{*}{Temperature [K]}} & \multicolumn{1}{l}{} &                                                                                                      \multicolumn{7}{c}{1T-MoSe$_2$}                                                                                                      \\ \cline{3-9}
			                  \multicolumn{1}{c}{}                   & \multicolumn{1}{l}{} &                                     \multicolumn{3}{c}{ Strain-$x$}                                      & \multicolumn{1}{l}{} &                                      \multicolumn{3}{c}{Strain-$y$}                                      \\ \cline{3-5}\cline{7-9}
			                  \multicolumn{1}{c}{}                   & \multicolumn{1}{l}{} & \multicolumn{1}{c}{$Y_M$ {[}GPa{]}} & \multicolumn{1}{c}{FS {[}\%{]}} & \multicolumn{1}{c}{US {[}GPa{]}} & \multicolumn{1}{l}{} & \multicolumn{1}{c}{$Y_M$ {[}GPa{]}} & \multicolumn{1}{c}{FS {[}\%{]}} & \multicolumn{1}{c}{US {[}GPa{]}} \\ \cline{1-1}\cline{3-5}\cline{7-9}
			                          10 K                            &                      &               127.02                &              12.03              &              11.20               &                      &               124.32                &              14.62              &              11.60               \\
			                \multicolumn{1}{c}{300 K}                 & \multicolumn{1}{l}{} &     \multicolumn{1}{c}{114.41}      &    \multicolumn{1}{c}{7.31}     &     \multicolumn{1}{c}{7.70}     & \multicolumn{1}{l}{} &     \multicolumn{1}{c}{111.80}      &    \multicolumn{1}{c}{8.28}     &     \multicolumn{1}{c}{8.02}
		\end{tabular}
		
		\begin{tabular}{clccclccc}
			\hline
			\multicolumn{1}{c}{\multirow{3}{*}{Temperature [K]}} & \multicolumn{1}{l}{} &                                                                                                      \multicolumn{7}{c}{1T-MoTe$_2$}                                                                                                      \\ \cline{3-9}
			                  \multicolumn{1}{c}{}                   & \multicolumn{1}{l}{} &                                     \multicolumn{3}{c}{ Strain-$x$}                                      & \multicolumn{1}{l}{} &                                      \multicolumn{3}{c}{Strain-$y$}                                      \\ \cline{3-5}\cline{7-9}
			                  \multicolumn{1}{c}{}                   & \multicolumn{1}{l}{} & \multicolumn{1}{c}{$Y_M$ {[}GPa{]}} & \multicolumn{1}{c}{FS {[}\%{]}} & \multicolumn{1}{c}{US {[}GPa{]}} & \multicolumn{1}{l}{} & \multicolumn{1}{c}{$Y_M$ {[}GPa{]}} & \multicolumn{1}{c}{FS {[}\%{]}} & \multicolumn{1}{c}{US {[}GPa{]}} \\ \cline{1-1}\cline{3-5}\cline{7-9}
			                          10 K                            &                      &               117.86                &              13.28              &              11.12               &                      &               114.40                &              16.06              &              11.29               \\
			                \multicolumn{1}{c}{300 K}                 & \multicolumn{1}{l}{} &     \multicolumn{1}{c}{107.16}      &    \multicolumn{1}{c}{8.03}     &     \multicolumn{1}{c}{7.91}     & \multicolumn{1}{l}{} &     \multicolumn{1}{c}{103.02}      &    \multicolumn{1}{c}{9.23}     &     \multicolumn{1}{c}{8.19}     \\ \hline
		\end{tabular}

		\caption{Elastic properties ($Y_{M}$, in units of GPa), fracture strain ($FS (\%)$), and the maximum stress $\sigma_{US}$ (which is called Ultimate Tensile Strength US (GPa)) that were obtained by fitting the stress-strain curves for the TMD monolayers investigated here.}
		\label{table:elasprop}
	\end{table*}

\section{Conclusions}

In summary, we carried out fully-atomistic reactive molecular dynamics simulations to perform a comparative study on the elastic properties and fracture patterns of MoX$_{2}$ (X=S,Se,Te) membranes, in the 2H and 1T phases, within the framework of the Stillinger-Weber potential. The results showed that the fracture mechanism of a 2H-MoS$_2$ monolayer occurs through an abrupt rupture followed by fast crack propagation. A different fracture trend is realized for a 1T-MoS$_2$ membrane. The fracture dynamics of this material leads to a brittle structure. Both MoSe$_{2}$ and MoTe$_{2}$ presented a fracture mechanism defined by a fast crack propagation followed by an abrupt rupture of the membranes into parts with a good degree of integrity (i.e., no brittle structures were observed). These results suggest that these monolayers may present greater structural stability than the MoS$_2$ ones. The highest Young's Modulus has obtained for 2H-MoSe($_2$) monolayer at 10K (154.65 GPa). The TMD of higher tensile strength is the 2H-MoSe$_{2}$ membrane at 10K (25.98 GPa). Generally, the critical strain to realize the TMD membranes fracture is considerably higher when the strain was applied along the $y$-direction. It was also obtained here as a general trend that the translated arrangement of the chalcogen atoms in the 1T phase can contribute to diminishing their structural stability when compared with TMD membranes in the 2H phase. As a consequence, among the 2H and 1T phases, the 1T presented lower tensile strength. 

\section*{Acknowledgements}
The authors gratefully acknowledge the financial support from Brazilian research agencies CNPq, FAPESP, and FAP-DF.  M.L.P.J. gratefully acknowledge the financial support from CAPES grant 88882.383674/2019-01. R.T.S.J. gratefully acknowledge, respectively, the financial support from CNPq grants 465741/2014-2 and 312180/2019-5, CAPES grant 88887.144009/2017-00, and FAP-DF grants 0193.001366/2016 and 0193.001365/2016. L.A.R.J acknowledges the financial support from a Brazilian Research Council FAP-DF and CNPq grants $00193.0000248/2019-32$ and $302236/2018-0$, respectively. L.A.R.J. and M.L.P.J. acknowledge CENAPAD-SP for providing the computational facilities.

\printcredits

\bibliographystyle{unsrt}

\bibliography{references}

\end{document}